\documentclass[12pt]{article}

\usepackage[height=8.85in,width=6.45in]{geometry}

\usepackage[utf8]{inputenc}
\usepackage{amsmath}
\usepackage{amssymb}
\usepackage{mathtools}
\numberwithin{equation}{section}
\usepackage{slashed}
\usepackage{braket}
\usepackage{hyperref}
\usepackage{wasysym}
\usepackage{eufrak}
\usepackage{graphicx}
\usepackage{color}
\allowdisplaybreaks[2]
\usepackage{slashed}
\usepackage{tikz}
\usepackage[compat=1.1.0]{tikz-feynman}
\usetikzlibrary{decorations.pathmorphing,decorations.markings}
\usepackage{subfig}
\usepackage{verbatim}
\usepackage{cite}
\usepackage{xcolor}
\usepackage{mdframed}

\renewenvironment{figure}[1][]{
  \begin{originalfigure}[#1]
    \begin{mdframed}[linecolor=black!0,backgroundcolor=black!0]
}{
    \end{mdframed}
  \end{originalfigure}
}

\def\Nequals#1{$\mathcal{N}{=}\,#1$}
\def\su{\mathfrak{su}}

\def\so{\mathfrak{so}}
\def\usp{\mathfrak{usp}}
\def\diag{\mathop{\mathrm{diag}}}
\def\vev#1{\langle{#1}\rangle}

\usepackage{allrunes}
\newcommand{\hconj}[1]{{#1}^\dagger}
\newcommand{\hconjp}[1]{\hconj{({#1})}}
\newcommand{\com}[2]{[\,{#1} \, , \,{#2} \,]}
\newcommand{\acom}[2]{\{\,{#1} \, , \,{#2} \,\}}

\newcommand{\rnum}{\mathbb{R}}

\newcommand{\mathrune}[1]{\text{\textarc{#1}}}

\begin{document}

\begin{titlepage}

\begin{flushright}
IPMU-17-0088
\end{flushright}

\vskip 4cm

\begin{center}

{\large\bfseries On  hydrogen-like bound states in \Nequals4 super Yang-Mills}

\vskip 1cm
Yusuke Sakata$^1$, Robin Schneider$^{1,2}$, Yuji Tachikawa$^1$ and Takemasa Yamaura$^1$
\vskip 1cm

\begin{tabular}{ll}
1.  & Kavli Institute for the Physics and Mathematics of the Universe, \\
& University of Tokyo,  Kashiwa, Chiba 277-8583, Japan\\
2. & Department of Physics and Astronomy, Uppsala University,\\
& Box 516, SE-75120 Uppsala, Sweden
\end{tabular}

\vskip 1cm

\end{center}

\noindent 
Using relativistic quantum mechanics,
we study the spectrum of a non-BPS two-particle bound state in the massive phase of \Nequals4 super Yang-Mills, in the limit  when one of the particles is infinitely heavier than the other.
We find that the spectrum shows the exact $n^2$ degeneracy for each principal quantum number $n$, just as in the strict non-relativistic limit.
This is in line with the findings of Caron-Huot and Henn, who studied the same system in the large $N$ limit with the technique of integrability and the dual conformal symmetry. 
%Their calculation was for arbitrary value of the coupling constant including the full effect of quantum field theory, and they commented on its implication on the dual conformal symmetry.

\end{titlepage}

\section{Introduction}
The quantum-mechanical spectrum of two particles bound by the Coulomb force
\begin{align}
\mathcal{H}= \frac{\vec{p}}{2 \mu} - \frac{\alpha}{r} \; ,
\end{align}
shows an $n^2$ degeneracy for each principal quantum number $n$.
This can be understood from the existence of an accidental conserved charge \cite{Pauli} which extends the rotational symmetry $\so(3)$ to $\so(4)$.
One way to see this is as follows. 
The states at a given principal quantum number $n$ transform as \begin{equation}
V_0\oplus V_1\oplus \cdots \oplus V_n
=V_{n/2} \otimes V_{n/2}
\end{equation} where $V_l$ is the spin-$l$ representation of $\so(3)$. 
The right hand side has a natural action of the extended $\so(4)\simeq \su(2) \times \su(2)$ symmetry, where the first $\su(2)$ acts on the first $V_{n/2}$ and the second acts on the second $V_{n/2}$.
For more details, see e.g.~the extensive review article \cite{Bander:1965rz} and references therein.
Of course this extended symmetry is broken in the real world, due to various corrections.
For a detailed account of these corrections, see e.g.~\cite{Eides:2000xc}.

Recently, it was shown by Caron-Huot and Henn \cite{Caron-Huot:2014gia} that in \Nequals4 super Yang-Mills theory, there is a two-particle bound state where  relativistic corrections still preserve the extended $\so(4)$ symmetry.
The analysis there was carried out using sophisticated techniques of integrability of the \Nequals4 theory, developed during the last decade and a half.

Caron-Huot and Henn interpreted their results as a manifestation of the dual conformal symmetry in the massive phase of the theory in the large $N$ limit.
Let us comment on this point. 
The \Nequals4 super Yang-Mills theory in the massless phase is long established to be symmetric under the superconformal group acting on the spacetime $x^\mu$ and its superpartners.
More recently, it is realized that the same theory in the large $N$ limit also has another, or dual,  superconformal symmetry, which, roughly speaking, acts on the momentum space $p^\mu$ and its superpartners, see e.g.~the textbook \cite{Elvang:2013cua} and references therein.
This dual superconformal symmetry has been generalized to some amplitudes in the massive phase \cite{Alday:2009zm,CaronHuot:2010rj,Dennen:2010dh,Plefka:2014fta}.

Turning back to the hydrogen atom, already in 1935, Fock realized that the $\so(4)$ symmetry of the hydrogen spectrum is a subgroup of the conformal transformation acting on the momentum vector $\vec p\in \rnum^3$ \cite{Bander:1965rz,Fock}.
More specifically, he mapped the momentum space $\rnum^3$ by an inverse stereographic projection to $S^3$, and showed that the Coulomb interaction written in this manner is invariant under the natural $\so(4)$ action on this $S^3$.
Therefore it is indeed natural to attribute this $\so(4)$ symmetry to be the unbroken part of the dual conformal symmetry $\so(4,2)$ in the presence of the rest mass of the bound state.

In this note, we consider the same bound state in the same \Nequals4 super Yang-Mills theory, from  a totally elementary point of view. 
Our analysis goes as follows.
First,  we consider the standard Klein-Gordon equation of a particle in the Coulomb potential, modified to include the effect of the massless scalar exchange. 
We show that it can be easily solved exactly, and that it preserves the $n^2$ degeneracy.
Second, we perform a schematic analysis of Feynman diagrams contributing to the two-particle bound states, and show that in the limit where one of the particles is infinitely heavier, 
the spectrum can  indeed be found by solving the modified Klein-Gordon equation discussed in the first part. 

Our analysis only scratches the surface of the work done by Caron-Huot and Henn: in their work  the same system was studied in the large $N$ limit at arbitrary coupling including full effects of quantum field theory, while in this note we only treat it at the level of relativistic quantum mechanics, without taking the large $N$ limit.
Our intention is purely educational: our elementary analysis would be understandable to anyone with a basic knowledge of quantum field theory, and would not require the mastery of the techniques of integrability of \Nequals4 super Yang-Mills. 
Hopefully this short note would serve as an introduction to this interesting work of Caron-Huot and Henn.

Before proceeding, let us mention that \Nequals1 and \Nequals2 supersymmetric analogues of hydrogen atoms have been considered since  the 80's \cite{Buchmuller:1981bp,DiVecchia:1985xm}.
More recently, the \Nequals1 version was revisited in \cite{Herzog:2009fw,Rube:2009yc}.
They have found that the split of the degeneracy due to the relativistic correction is milder with more supersymmetry.
The result of \cite{Caron-Huot:2014gia} and ours can be thought of as a confirmation of the continuation of this trend to \Nequals4 supersymmetry.\footnote{%
We also note that there is an approach trying to find a supersymmetric structure in the relativistic Hamiltonian of the standard hydrogen atom \cite{Katsura:2004pw}.}

The rest of the note is organized as follows.
In Sec.~\ref{relQM}, we study the Klein-Gordon equation of a particle coupled to both the Coulomb field and to the classical massless scalar field, which captures the main modification due to the \Nequals4 supersymmetry. The solution shows the $n^2$ degeneracy.
In Sec.~\ref{setup}, we  show that this modified Klein-Gordon equation does arise from the \Nequals4 super Yang-Mills.
In Sec.~\ref{other}, we briefly comment on further corrections to the spectrum.
Finally in Sec.~\ref{outlook}, we conclude with a brief discussion of further directions of research.

\section{Relativistic quantum-mechanical analysis}
\label{relQM}

Let us first consider the Klein-Gordon equation of a charged particle in a Coulomb field.
It is given by \begin{equation}
\left[-(\partial_t - iA_t)^2 + \triangle -m^2 \right]\phi=0, \qquad A_t=\alpha/r.
\end{equation}  
The energy eigenvalues can be obtained by the standard separation of variables (see e.g. ~Chap.~21 of \cite{BetheTextbook}); one obtains \begin{equation}
E_{n,l}=m \left[ 1+\left(\frac{\alpha}{n+\sqrt{(l+1/2)^2-\alpha^2}-(l+1/2)} \right)^2\right]^{-1/2}.
\end{equation}
In the non-relativistic limit, we have \begin{equation}
E_{n,l}=m\left[ 1-\frac{\alpha^2}{2n^2}-\frac{\alpha^4}{2n^4} \bigg( \frac{n}{l+1/2}-\frac34 \bigg)+\cdots\right]
\end{equation}
which clearly shows the relativistic splitting of the $n^2$ degeneracy.

The main effect of having \Nequals4 supersymmetry is that i) the mass of a charged particle is given by a vacuum expectation value (vev) of another scalar $\tilde\phi$ which is a superpartner of the photon, and that ii) this massless scalar is also exchanged between the charged particles.
 When we consider two particles of like charges, the attractive potential coming from the massless scalar exactly cancels the repulsive potential coming from the photon. 
Here we consider two particles of unlike charges, for which the Klein-Gordon equation now reads
\begin{equation}
\left[-(\partial_t -iA_t)^2 + \triangle - \tilde\phi^2 \right]\phi=0, \qquad A_t=\frac{\alpha}r,\quad \tilde\phi=m-\frac{\alpha}r\label{modifiedKG}
\end{equation}
where the form of $\tilde\phi$ is chosen so that the particle $\phi$ has mass $m$ and the leading $1/r$ potential exactly cancels in the case of like charges.
A slight rewrite using  $\partial_t=-iE$  gives \begin{equation}
\left[(m+E) (m-E-\frac{2\alpha}{r}) - \triangle \right]\phi=0
\end{equation} whose eigenvalues are easily seen to be \begin{equation}
E_{n,l}=m\left[1-\frac{2\alpha^2}{n^2+\alpha^2}\right]
=m\left[1-\frac{2\alpha^2}{n^2}+\frac{2\alpha^4}{n^4}+\cdots\right]
\end{equation}
which indeed shows the complete $n^2$ degeneracy kept intact.

\section{Extracting the Klein-Gordon equation}
\label{setup}
Now we would like to show that the modified Klein-Gordon equation \eqref{modifiedKG} actually governs the spectrum of a bound state in the \Nequals4 super Yang-Mills in a suitable limit.

The Lagrangian of the \Nequals4 $U(N)$ Yang-Mills was originally constructed in \cite{Brink:1976bc} and 
its bosonic part  has the following well-known form \begin{equation}
	\label{eq: N=4}
	\begin{split}
	\mathcal{L } &= \text{tr} \bigg\{
	-\frac{1}{4} F_{\mu\nu} F^{\mu\nu} 
	- \frac{1}{2} D_\mu \Phi_{i} D^\mu \Phi^{i} + \frac{g^{2}}{4} \sum_{i,j=1}^{6}  [\Phi_{i}, \Phi_{j} ]^{2} \bigg\} \; 
	\end{split}
\end{equation}
where each field is an $N\times N$ matrix and we use the convention that the coupling constant enters in the definition of the covariant derivative. 

We give a nonzero vev to $\Phi_1$ of the form \begin{equation}
\vev{\Phi_1}=\diag(m_1,\ldots,m_N)\label{vev}
\end{equation} while we  keep all other scalars $\Phi_{2,\ldots,6}$ to be zero.
The $(i,j)$-component of the fields obtains the mass $\propto |m_i-m_j|$,
while the diagonal components remain massless.
We can take $m_1:m_2:m_3 \sim 0:1:2000$ say, and call the $(1,2)$ component to be the `electron' while $(3,1)$ component to be the `proton'.
In the following we use the notation \begin{equation}
m_\text{e} := |m_1-m_2|, \qquad m_\text{p}:=|m_1-m_3|.
\end{equation}
We are interested in the limit $m_\text{e}/m_\text{p}\to 0$.

Due to the cancellation of the exchanges of the massless scalar and vector, the `electron' and the `anti-proton' do not bind, while the `electron' and the `proton' form a bound state.
This is the bound state we want to analyze.

The vev \eqref{vev} breaks the $\su(4)\simeq \so(6)$ R-symmetry to $\usp(4)\simeq \so(5)$.
The massive (anti-)BPS vector multiplets have the following content 
\begin{equation}
V_\text{BPS} \simeq  V_\text{anti-BPS} \simeq (\mathbf{1},\mathbf{5})_\phi \oplus (\mathbf{2},\mathbf{4})_\psi \oplus (\mathbf{3},\mathbf{1})_A.\label{(anti)BPS}
\end{equation}
under the little group $\so(3)$ times the unbroken R-symmetry group $\so(5)$.
Then, the states with the principal quantum number $n$ and the orbital angular momentum $l$ transform as \begin{equation}
V_l \otimes V_\text{BPS}\otimes V_\text{anti-BPS}
\label{massive-non-BPS}
\end{equation} 
where $V_l$ is the spin-$l$ representation of $\so(3)$.
For more details,  see Appendix~\ref{susy}, where we review the structure of general massive non-BPS states of \Nequals4 supersymmetry.

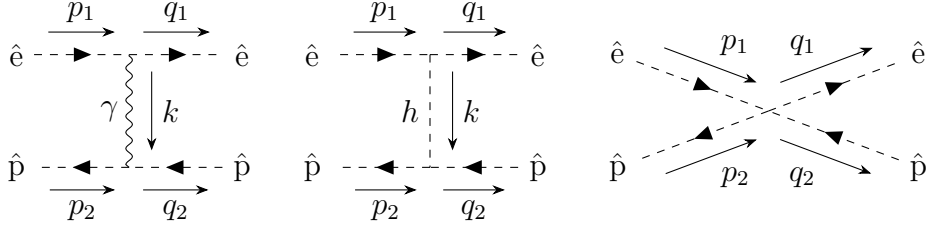
\begin{figure}[t]
	\centering
	\begin{tikzpicture}
	\begin{feynman}[scale =1.0]
	\vertex (a) at (-7,-3) {\( \text{\^{e}} \)};
	\vertex (b) at (-5.5,-3);
	\vertex (c) at (-4,-3) {\( \text{\^{e}} \)};
	\vertex (d) at (-5.5,-4.5) ;
	\vertex (e) at (-7,-4.5) {\( \text{\^{p}} \)};
	\vertex (f) at (-4,-4.5) {\( \text{\^{p}} \)};
	\diagram* {
		(a) -- [charged scalar, momentum=\(p_{1}\)] (b),
		(b) -- [charged scalar, momentum=\(q_{1}\)] (c),
		(d) -- [charged scalar, rmomentum=\(p_{2}\)] (e),
		(f) -- [charged scalar, rmomentum=\(q_{2}\)] (d),
		(b) -- [boson, edge label'=\(\gamma\), momentum=\(k\)] (d),
	};
	\end{feynman}
	\begin{feynman}[scale =1.0]
	\vertex (a) at (-3,-3) {\( \text{\^{e}}\ \)};
	\vertex (b) at (-1.5,-3);
	\vertex (c) at (0,-3) {\( \text{\^{e}}\ \)};
	\vertex (d) at (-1.5,-4.5) ;
	\vertex (e) at (-3,-4.5) {\( \text{\^{p}}\ \)};
	\vertex (f) at (0,-4.5) {\( \text{\^{p}}\ \)};
	\diagram* {
		(a) -- [charged scalar, momentum=\(p_{1}\)] (b),
		(b) -- [charged scalar, momentum=\(q_{1}\)] (c),
		(d) -- [charged scalar, rmomentum=\(p_{2}\)] (e),
		(f) -- [charged scalar, rmomentum=\(q_{2}\)] (d),
		(b) -- [scalar, edge label'=\(h\), momentum=\(k\)] (d),
	};
	\end{feynman}
	\begin{feynman}[scale =1.0]
	\vertex (a) at (1,-3) {\( \text{\^{e}}_{} \)};
	\vertex (b) at (3,-3.75);
	\vertex (c) at (5,-3) {\(\text{\^{e}}_{} \)};
	\vertex (d) at (1,-4.5) {\(\text{\^{p}}_{}\)};
	\vertex (e) at (5,-4.5) {\(\text{\^{p}}_{} \)};
	\diagram* {
		(a) -- [charged scalar, momentum=\(p_{1}\)] (b),
		(b) -- [charged scalar, momentum=\(q_{1}\)] (c),
		(b) -- [charged scalar, rmomentum=\(p_{2}\)] (d),
		(e) -- [charged scalar, rmomentum=\(q_{2}\)] (b),
	};
	\end{feynman}
	\end{tikzpicture}
	\caption{The three tree level diagrams of \^{e} \^{p} $ \rightarrow $ \^{e}  \^{p} scattering.}
	\label{fig: sbosonbound}	
\end{figure}

We note that in the massive non-BPS multiplet \eqref{massive-non-BPS},
the symmetric-traceless representation $\mathbf{14}$ of $\so(5)$ appears exactly once, in the scalar-scalar bound state.
Since every state in this non-BPS multiplet has the same energy, 
we can just study the energy of the bound state in this representation $\mathbf{14}$.
This allows us to restrict the two particles in the bound state to be the scalar component within the (anti-)BPS multiplets.
The processes which contribute to the bound state in this $\so(5)$ representation are shown in Fig.~\ref{fig: sbosonbound},
where we denoted the `scalar electron' by \^e and the `scalar proton' by \^p.

The corresponding amplitudes to the leading non-relativistic order are given as follows.
Both the \emph{first} amplitude corresponding to a photon exchange 
and the \emph{second} amplitude corresponding to a massless scalar exchange are given by
\begin{equation}
\label{eq: scatteringmain}
i \mathcal{M}_{1}  \simeq i\mathcal{M}_2 \simeq \frac{4 m_{\text{e}} m_{\text{p}} g^{2}}{\vec k^{2}} 
\end{equation}
where $\vec k$ is the momentum transfer.
The first two amplitudes are insensitive to the R-symmetry representations,
while the \emph{third} one from the direct four-particle interaction depends on it. 
In the representation $\mathbf{14}$ we find that it is given by \begin{equation}
i \mathcal{M}_{3} =-2 g^{2} . \; 
\end{equation}

In the relativistic quantum field theory the single-particle states are usually normalized so that $\vev{p|q}=2\sqrt{(m^2+\vec p^2)}\delta^3(\vec p-\vec q)$ while in the non-relativistic quantum mechanics it is normalized so that $\vev{\vec p|\vec q}=\delta^3(\vec p-\vec q)$. 
Therefore in the non-relativistic normalization the total scattering amplitude is given by 
\begin{align}
\mathcal{M}_{\text{NR}} \simeq \frac{\mathcal{M}}{ 4 m_\text{e} m_\text{p}} \; .
\end{align}

This means that in the limit $m_\text{e}/m_\text{p}\to 0$, the direct delta-function interaction term $i\mathcal{M}_3$ drops out,
and the effect of the infinitely heavy particle can be replaced by the $1/r$ potential created by the massless vector and the massless scalar.
Defining $\alpha$ by the requirement $A_t=\alpha/r$, the value of the massless scalar background is of the general form $\tilde\phi(x)=cm-c'\alpha/r$ where $c$ and $c'$ are numerical coefficients.
They can be fixed by a computation from the original Lagrangian \eqref{eq: N=4} or just by demanding that the field $\phi$ to have mass $m$ and the leading $1/r$ potential to exactly cancel in the case of like charges. 
Taking the latter approach immediately  gives $c=c'=1$.
In conclusion, we find that the energy spectrum of the bound state, in this limit, can be obtained by just solving the modified Klein-Gordon equation \eqref{modifiedKG}. 
This is what we wanted to show in this section.

\section{Comments on further corrections}
\label{other}

In this paper we only considered  corrections to the spectrum at the level of the relativistic quantum mechanics in the limit where one of the particle is infinitely heavier than the other.
Here we briefly discuss further corrections from various sources. 

\paragraph{Hyperfine splitting and recoil effect:}
For the real-world hydrogen the leading relativistic correction scaling with $\mu \alpha^4$ is called the fine structure, and the corrections of the form $ \mu \alpha^4 m_\text{e}/m_\text{p}$ are called the hyperfine splitting or the recoil effects.
The hyperfine splitting is the one depending on the spin of the heavier particle, and the recoil effect is the one independent of it.
Note that they vanish in the limit $m_\text{e}/m_\text{p}\to 0$.
 
So far we showed that the fine splitting is absent in the bound state under our considerations.
There is no hyperfine splitting, from the simple reason that the bound state form a single non-BPS multiplet under the supersymmetry.
It would be nice to extend our study to the recoil effect, by making $m_\text{e}/m_\text{p}$ finite.

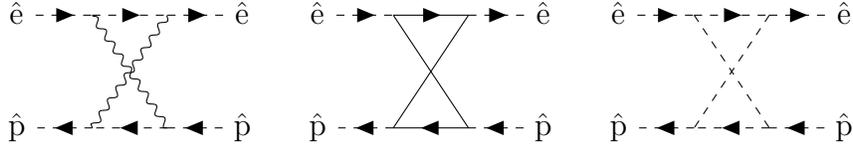
\begin{figure}[t]
	\centering
	\begin{tikzpicture}
	\begin{feynman}[scale =1.0]
	\vertex (a) at (-7,-3) {\( \text{\^{e}}_{} \)};
	\vertex (b) at (-6,-3);
	\vertex (g) at (-5,-3);
	\vertex (c) at (-4,-3) {\( \text{\^{e}}_{} \)};
	\vertex (d) at (-6,-4.5) ;
	\vertex (h) at (-5,-4.5) ;
	\vertex (e) at (-7,-4.5) {\( \text{\^{p}}_{} \)};
	\vertex (f) at (-4,-4.5) {\( \text{\^{p}}_{} \)};
	\diagram* {
		(a) -- [charged scalar] (b),
		(b) -- [charged scalar] (g),
		(g) -- [charged scalar] (c),
		(d) -- [charged scalar] (e),
		(h) -- [charged scalar] (d),
		(f) -- [charged scalar] (h),
		(b) -- [boson] (h),
		(g) -- [boson] (d),
	};
	\end{feynman}
	\begin{feynman}[scale =1.0]
	\vertex (a) at (-3,-3) {\( \text{\^{e}}_{} \)};
	\vertex (b) at (-2,-3);
	\vertex (g) at (-1,-3);
	\vertex (c) at (0,-3) {\( \text{\^{e}}_{} \)};
	\vertex (d) at (-2,-4.5) ;
	\vertex (h) at (-1,-4.5) ;
	\vertex (e) at (-3,-4.5) {\( \text{\^{p}}_{} \)};
	\vertex (f) at (0,-4.5) {\( \text{\^{p}}_{} \)};
	\diagram* {
		(a) -- [charged scalar] (b),
		(b) -- [fermion] (g),
		(g) -- [charged scalar] (c),
		(d) -- [charged scalar] (e),
		(h) -- [fermion] (d),
		(f) -- [charged scalar] (h),
		(b) -- [plain] (h),
		(g) -- [plain] (d),
	};
	\end{feynman}
	\begin{feynman}[scale =1.0]
	\vertex (a) at (1,-3) {\( \text{\^{e}}_{} \)};
	\vertex (b) at (2,-3);
	\vertex (g) at (3,-3);
	\vertex (c) at (4,-3) {\( \text{\^{e}}_{} \)};
	\vertex (d) at (2,-4.5) ;
	\vertex (h) at (3,-4.5) ;
	\vertex (e) at (1,-4.5) {\( \text{\^{p}}_{} \)};
	\vertex (f) at (4,-4.5) {\( \text{\^{p}}_{} \)};
	\diagram* {
		(a) -- [charged scalar] (b),
		(b) -- [charged scalar] (g),
		(g) -- [charged scalar] (c),
		(d) -- [charged scalar] (e),
		(h) -- [charged scalar] (d),
		(f) -- [charged scalar] (h),
		(b) -- [scalar] (h),
		(g) -- [scalar] (d),
	};
	\end{feynman}
	\end{tikzpicture}
	\caption{Examples of crossed one-loop diagrams which can also contribute to the bound state energy  at $\alpha^4$.}
	\label{fig: crossed}	
\end{figure}

In principle we can evaluate the Feynman diagrams such as those shown in Figure~\ref{fig: sbosonbound} to the required order and plug it in to the Bethe-Salpeter equation determining the bound state spectrum.
However, at the same order $\alpha^4$, we also need to include the contributions from crossed ladder diagrams such as shown in Figure~\ref{fig: crossed}.

The basic reason is as follows. 
To the Hamiltonian of the bound state, the tree-level diagram contributes a term of the form $\sim\alpha /r + \cdots$ while the one-loop diagrams contribute terms of the form $\sim\alpha^2/r^2 + \cdots$.
Recall now that the expectation value of the operator $1/r^n$ in the standard wavefunctions of the non-relativistic hydrogen atom scales as $\sim \alpha^n$. 
Therefore, the leading contribution from the tree-level diagram is $\sim \alpha^2$ as it should be, with a subleading $\alpha^4$ term from the relativistic kinematics,
and the leading contribution from the one-loop diagram is also of order $\sim\alpha^4$.
The straight ladder diagrams are two-particle irreducible and therefore do not contribute in the Bethe-Salpeter equation.
%The question still remains which of the crossed one-loop diagrams has a non zero contribution at $\alpha^4$.

It is known \cite{lindgren1990gauge} that the crossed photon diagram, the first of the three shown in Figure~\ref{fig: crossed}, does not contribute in the case of QED in the Coulomb gauge. 
But there are additional crossed diagrams in our case, including the crossed gaugino, the crossed scalar, and many others, and we suspect that they do indeed have a nonzero contribution at this order.
To say anything definite, we need to have a more detailed study of the system.

\paragraph{Lamb shift:}
In the case of real-world hydrogen,
the leading radiative correction is the famous Lamb shift and is of order $\alpha^5 \log \alpha$.
In the supersymmetric case, the radiative gaugino and photon corrections are again of order $\alpha^5\log \alpha$ \cite{Rube:2009yc}.

However, the radiative correction from the massless scalar is known to give a correction of order $\alpha^3\log\alpha$  \cite{Pineda:2007kz}, as reproduced by the integrability technique in \cite{Caron-Huot:2014gia}.
Note that this is parametrically larger than the relativistic quantum-mechanical corrections, which are of order $\alpha^4$.
Therefore, strictly speaking, it is of no use to isolate the relativistic quantum-mechanical corrections in the present case, as we did in this paper.

Here we give a rough argument that this effect produces no $l$ dependence; see \cite{Caron-Huot:2014gia} for details.
Note that in the end any computation of the correction boils down to an evaluation of the expectation value of the operator $\Delta H=\alpha^n V(\vec r,\vec p)$ modifying the non-relativistic Hamiltonian in the standard uncorrected hydrogen wavefunction. 
A one-loop correction corresponds to $\alpha^2$ contributions.
For the photon exchange, we know $V\sim \vec p\,^2/r $ or $\sim \delta^3(\vec r)$,
whose expectation values give $(\mu\alpha)^3$, in total giving a correction of order $\alpha^{2+3}$.
In general, the expectation value of a dimension-$d$ operator is of order $(\mu\alpha)^d$.
To produce an $\alpha^3$ shift, the operator $V$ should be of dimension 1, which restricts its form to be $\sim 1/r$, whose expectation value is of the form $\sim (\mu\alpha)/n^2$, and cannot produce an $l$-dependent contribution. 
The subleading correction from the massless scalar would be of order $\alpha^5 \log \alpha$ as in the standard Lamb shift.

\paragraph{Mixing and decays:}
The vev \eqref{vev} breaks $U(N)$  to $U(1)^N$, and we have been considering the bound state of a particle with  gauge charge $(q_1,q_2)=(1,-1)$ and another with $(q_1,q_3)=(-1,1)$, with the total charge $(q_2,q_3)=(-1,1)$.
There are many additional states with the total charge $(q_2,q_3)=(-1,1)$ which can mix with our bound state.

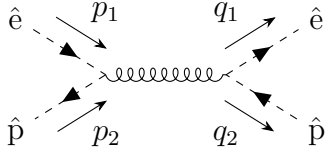
\begin{figure}[t]
	\centering
	\begin{tikzpicture}
	\begin{feynman}[scale =1.0]
	\vertex (a) at (1,-3) {\( \text{\^{e}} \)};
	\vertex (b) at (2.2,-3.75);
	\vertex (c) at (5,-3) {\(\text{\^{e}} \)};
	\vertex (bb) at (3.8,-3.75);
	\vertex (d) at (1,-4.5) {\(\text{\^{p}}\)};
	\vertex (e) at (5,-4.5) {\(\text{\^{p}} \)};
	\diagram* {
		(a) -- [charged scalar, momentum=\(p_{1}\)] (b),
		(bb) -- [charged scalar, momentum=\(q_{1}\)] (c),
		(b) -- [gluon] (bb),
		(b) -- [charged scalar, rmomentum=\(p_{2}\)] (d),
		(e) -- [charged scalar, rmomentum=\(q_{2}\)] (bb),
	};
	\end{feynman}
	\end{tikzpicture}
	\caption{Another possible contribution at $\alpha^4$.}
	\label{fig: annih}	
\end{figure}

The first possibility is that we just have a single particle with the total charge $(q_2,q_3)=(-1,1)$ in the intermediate state.
The corresponding Feynman diagram is shown in figure~\ref{fig: annih}.
The diagram itself is of order $\alpha$, and the process proceeds via two constituent particles merging into one, therefore with the delta function interaction $\delta^3(\vec r)$.
In total the correction is of order $\alpha^{1+3}$.
In the case of the positronium in the real world, this indeed contributes to the splitting of the ortho-positronium and the para-positronium at this order.
In our case, however, the contribution is zero, since the intermediate channel is a massive (anti)-BPS state and there is no component with the representation $\mathbf{14}$ to mix with our states.

Next, we can have, in the intermediate state, massive non-BPS multiplets which are composed of the $(n,2)$-component and the $(3,n)$-component of the gauge multiplet, where $n\neq 1$.
We can safely neglect the cases when $n\neq 2,3$ by taking $m_{n\neq 1,2,3}$ very large, since these will then have masses of order $2m_n-m_2-m_3 \gg m_\text{e}+m_\text{p}$.

When $n=2$ or $=3$, this is  a decay channel where \^e and \^p merge to form a massive (anti)-BPS state with charge $(q_2,q_3)=(-1,1)$, simultaneously emitting a massless vector multiplet. 
The diagram itself is of order $\alpha$, and again this proceeds via a point interaction. 
Therefore the decay width is of order $\alpha^2 \vev{\delta^3(\vec r)} \sim \alpha^5$.

\section{Discussions}
\label{outlook}
In this paper we studied the  relativistic correction to the spectrum of a Coulombic bound state of two particles in the massive phase of \Nequals4 super Yang-Mills, using relativistic quantum mechanics.
We found that in the $m_\text{e}/m_\text{p}\to 0$ limit, the correction keeps the $n^2$ degeneracy, meaning that it preserves the extended $\so(4)$ invariance.

As for future directions, firstly, we would like to analyze the spectrum in the finite $m_\text{e}/m_\text{p}$ case.
An approach would be to extend the pNRQED framework \cite{Pineda:1997bj,Pineda:1997ie} to include the presence of the massless scalar, as in \cite{Pineda:2007kz}.

Secondly, it would be interesting to understand more fully the structure of the dual conformal symmetry in the massive phase.
The states at a given principal quantum number $n$ transform as \begin{equation}
(V_0\oplus V_1\oplus \cdots \oplus V_n) \otimes V_\text{BPS}\otimes V_\text{antiBPS}
=V_{n/2} \otimes V_{n/2} \otimes V_\text{BPS}\otimes V_\text{antiBPS}.
\end{equation} 
One immediate question is how the extended $\so(4)\simeq \su(2)\times \su(2)$ should act on the BPS and anti-BPS linear combinations of the supercharges. 
It is tempting to think that one $\su(2)$ acts on $V_\text{BPS}$ and another acts on $V_\text{antiBPS}$.

The analyses given in the literature so far typically use an extension to higher dimensions to regard massive states in four dimensions as massless states in higher dimensions. 
The modified Klein-Gordon equation \eqref{modifiedKG} indeed has this structure;
it would be interesting to supersymmetrize it and and study its symmetry group.
Also, considering the importance of the central charge $Z^{[ij]}$ which modifies the anticommutator of two supersymmetry generators in the massive phase, 
another approach would be to try to add this central charge, in some way or other, to the Yangian symmetry which combines both the original and the dual superconformal symmetry \cite{Drummond:2009fd}, possibly following the discussions in \cite{Plefka:2014fta}.

We would like to come back to these questions in the future.

\section*{Acknowledgements}
The authors thank S. Caron-Huot and C. P. Herzog for discussions.
YS and TY are partially supported by the Programs for Leading Graduate Schools, MEXT, Japan, via the Leading Graduate Course for Frontiers of Mathematical Sciences and Physics. 
RS is partially supported by the Jasso scholarship from the Graduate School of Science, the University of Tokyo.
YT is in part supported by JSPS KAKENHI Grant-in-Aid (Wakate-A), No.17H04837 
and JSPS KAKENHI Grant-in-Aid (Kiban-S), No.16H06335, and 
also supported in part by WPI Initiative, MEXT, Japan at IPMU, the University of Tokyo.

\appendix

\section{Structure of massive non-BPS bound state}
\label{susy}
The structure of representations of non-conformal supersymmetries in general dimensions were worked out in \cite{Nahm:1977tg,Strathdee:1986jr}.
Here we give a brief account for the particular case of \Nequals4 supersymmetry in four dimensions, in the special case where the central charge breaks $\su(4)\simeq \so(6)$ R-symmetry to $\usp(4)\simeq \so(5)$.

Let us say that the four-momentum of the total system is $(m,0,0,0)$ with the little group $\so(3)$ and the central charge is $Z^{IJ}=zJ^{IJ}$ where $J^{IJ}$ is the invariant tensor of the $\usp(4)$ with $z\in \rnum$.
We would like to understand  the structure of the irreducible representation of the little group $\so(3)$, the R-symmetry $\so(5)$ together with
the supersymmetry generators whose  anticommutation relations are
	\begin{align}
		\acom{Q_a^I}{Q_b^J} &= \epsilon_{ab} zJ^{IJ}\, , & 
		\acom{Q_a^I}{\hconjp{Q_b^J}} &= 2m\delta_{ab}\delta^I_J\, .
	\end{align}
Here and in the following, the indices $a,b$ are for the little group $\so(3)$ and therefore there is no distinction of the dotted and undotted spinor indices. 

It is convenient to take the following linear combination of the supercharges instead: \begin{align}
	\mathrune{R}^{Ia}_\pm &:= \frac{1}{\sqrt{2}}( Q^{Ia} \pm J^{IJ}\epsilon^{ab}\hconjp{Q}_{Jb} )
\end{align} satisfying the reality condition \begin{equation}
		\hconjp{\mathrune{R}^{Ia}_\pm} = \overline{J}_{IJ}\epsilon_{ab} \mathrune{R}^{Jb}_\pm 
\end{equation} and the anticommutation relations \begin{align}
	\acom{\mathrune{R}^{Ia}_\pm}{\mathrune{R}^{Jb}_\pm} & = -(2m\pm z)J^{IJ}\epsilon^{ab}, &
	\acom{\mathrune{R}^{Ia}_\pm}{\mathrune{R}^{Jb}_\mp} & = 0.
\end{align}
In this form it is clear that we need the condition $2m\ge |z|$. 
Specializing to the non-BPS case, we assume $2m>|z|$ in the following.

By regarding the pair $(I,a)$ as a single index $\mu$ running from $1$ to $8$, we can make a further rescaling of the operators to make them in the form \begin{align}
			\acom{\gamma^\mu_\pm}{\gamma^\nu_\pm} &= 2\delta^{\mu\nu}, &
			\acom{\gamma^\mu_\pm}{\gamma^\nu_\mp} &= 0.
\end{align}
The $\so(3)\times \so(5)$ rotation is inside the $\so(8)$ rotation acting on the indices $\mu,\nu$. So in the rest of this summary we consider the representation of $\so(8)$ rotation $M^{\mu\nu}$ together with $\gamma^\mu_\pm$.

The operators
	\begin{equation}
		S^{\mu\nu}_\pm = \frac{1}{4} \com{\gamma^\mu_\pm}{\gamma^\nu_\pm}
	\end{equation}
satisfy the following relations: \begin{align}
	\com{S^{\mu\nu}_\pm}{\gamma^\rho_\pm}  &= \com{M^{\mu\nu}}{\gamma^\rho_\pm} , &
	\com{S^{\mu\nu}_\pm}{S^{\rho\sigma}_\pm}  &= \com{M^{\mu\nu}}{S^{\rho\sigma}_\pm}.
\end{align} 
Therefore, the operators \begin{equation}
T^{\mu\nu} := M^{\mu\nu}-S^{\mu\nu}_+ - S^{\mu\nu}_-
\end{equation}  satisfy the $\so(8)$ commutation relations, and  commute with $\gamma^\mu_\pm$'s.
Furthermore, two sets of operators $\{M^{\mu\nu}, \gamma^\mu_\pm\}$ and $\{T^{\mu\nu},\gamma^\mu_\pm\}$ generate the same algebra. 
Those who know the coset construction in 2d conformal field theories would recognize this as a baby version of it.

The structure of the irreducible representation of the latter is now clear: it is of the form $V\otimes V_\text{BPS}\otimes V_\text{anti-BPS}$, where $V$ is an irreducible representation of $\so(8)$ generated by $T^{\mu\nu}$ and 
$V_\text{(anti-)BPS}$ is the irreducible representation of $\gamma^\mu_\pm$ isomorphic to the \Nequals4 (anti-)BPS states.
$M^{\mu\nu}$ then acts as a diagonal subgroup.

\bibliographystyle{ytphys}
%\baselineskip=0.95\baselineskip%\let\ttfamily\relax
\let\bbb\bibitem\def\bibitem{\itemsep1pt\bbb}
\bibliography{ref}

\end{document}